\documentclass[9pt,conference]{IEEEtran}
\usepackage{amssymb,amsthm,amsmath,array}
\usepackage{graphicx}
\usepackage[caption=false,font=footnotesize]{subfig}
\usepackage{xspace}
\usepackage[sort&compress, numbers]{natbib}
\usepackage{stmaryrd}
\usepackage{xcolor}
\usepackage{mathtools}
\usepackage{float}
\usepackage{textcomp}
\usepackage{bbm}

\def\SNR{\mathrm{SNR}}
\def\R{\mathbb{R}}          									 	          

\newcommand{\lb}{\mathrm{\mathbf{l}}}

\renewcommand{\sb}{\mathrm{\mathbf{s}}}

\newcommand{\xb}{\mathrm{\mathbf{x}}}
\newcommand{\yb}{\mathrm{\mathbf{y}}}

\newcommand{\Db}{\mathrm{\mathbf{D}}}

\newcommand{\Lb}{\mathrm{\mathbf{L}}}

\newcommand{\Pb}{\mathrm{\mathbf{P}}}

\newcommand{\Rb}{\mathrm{\mathbf{R}}}
\newcommand{\Sb}{\mathrm{\mathbf{S}}}

\newcommand{\Vb}{\mathrm{\mathbf{V}}}

\newcommand{\gammab}{\mathrm{\bm{\gamma}}}

\newcommand{\rhob}{\mathrm{\bm{\rho}}}

\newcommand{\taub}{\mathrm{\bm{\tau}}}

\newcommand{\phib}{\mathrm{\bm{\phi}}}
\newcommand{\psib}{\mathrm{\bm{\psi}}}

\newcommand{\var}{\mathrm{Var}}

\newcommand{\Cc}{\mathcal{C}}

\usepackage{tcolorbox}
\usepackage{mdframed}
\usepackage{lipsum}

\newcommand{\eqdef}{\ensuremath{\stackrel{\mbox{\upshape\tiny def.}}{=}}}

\usepackage{amsmath}

\usepackage{amsthm}
\usepackage{amssymb}
\usepackage{amsfonts}
\usepackage{dsfont}
\usepackage{accents}
\usepackage{bm}
\usepackage{hyperref}

\usepackage{tabularx} 
\usepackage{caption}
\usepackage{multirow}
\usepackage{booktabs}
\usepackage{graphicx}
\graphicspath{{Figures/}}
\usepackage{tcolorbox}
\usepackage{wasysym}
\usepackage{longtable}
\usepackage{booktabs}

\usepackage{tikz,pgfplots}
\usepgfplotslibrary{groupplots}
\usetikzlibrary{intersections,calc,arrows,matrix,spy}
\usepackage{color}
\definecolor{darkred}{rgb}{0.6,0,0}
\definecolor{darkgreen}{rgb}{0,0.5,0}
\definecolor{darkblue}{rgb}{0,0,0.5}
\definecolor{SkyBlue}{rgb}{0.53, 0.81, 0.92}
\pgfplotsset{compat=1.5.1}

\usepackage{algorithm}
\usepackage{algpseudocode}
\usepackage[squaren,Gray]{SIunits}

\usepackage{placeins}
\usepackage{float}
\usepackage[normalem]{ulem}

\definecolor{mycolor1}{RGB}{47, 89, 245}%
\definecolor{mycolor2}{RGB}{214, 101, 41}%
\definecolor{mycolor3}{RGB}{19, 161, 11}%
\definecolor{mycolor4}{rgb}{0.,0.,1}%

\definecolor{sid}{RGB}{225, 0, 100}

\usepackage{graphicx}
\begin{document}
\title{Implicit Reconstructions from Deformed Projections for CryoET}
\author{\IEEEauthorblockN{
        Vinith Kishore\IEEEauthorrefmark{1}, 
        Valentin Debarnot\IEEEauthorrefmark{1}, and 
        Ivan Dokmani\'{c}\IEEEauthorrefmark{1}\IEEEauthorrefmark{2}
    }
    \IEEEauthorblockA{
        \IEEEauthorrefmark{1} University of Basel.\\
        \IEEEauthorrefmark{2} University of Illinois at Urbana-Champaign.}
}
\maketitle
\begin{abstract}
    Cryo-electron tomography (cryoET) is a technique that captures images of biological samples at different tilts, preserving their native state as much as possible. Along with the partial tilt series and noise, one of the major challenges in estimating the accurate 3D structure of the sample is the deformations in the images incurred during the acquisition. We model these deformations as continuous operators and estimate the unknown 3D volume using implicit neural representations. This framework allows to easily incorporate the deformation and estimate jointly the deformation parameters and the volume using a standard optimization algorithm. This approach doesn't require training data and can benefit from standard prior in the optimization procedure.
\end{abstract}

\section{Introduction}
Cryo-Electron Tomography (cryoET) \cite{doerr2017cryo} is an emerging imaging modality used to image biological samples such as cells, tissue, etc. These samples are flash-frozen and imaged using an electron microscope. The process is very similar to cryo Electron Microscopy (cryoEM) which has gained much attention due to its ability to resolve larger bio-molecules on a nanometer scale. Unlike cryoEM, cryoET is used for larger scale samples, where we can almost capture the interaction between the components in situ. In cryoET multiple images of the sample are acquired by tilting the thin frozen sample along its axis, at a discrete set of tilt angles.  Unlike cryoEM, we have a limited number of images to recover the complete information of the structure, this is known as the missing wedge in the Fourier domain, thus making the problem ill-possed. The tilt-series obtained from an unknown volume is defined as: 
\begin{equation}\label{eq:formation_model}
    \yb_m = \boldsymbol{P}(\boldsymbol{R}_{\theta_m}(\boldsymbol{\rho})) + \boldsymbol{\epsilon}_m
\end{equation}
where $\boldsymbol{\rho} \in \mathbbm{R}^{N \times N \times N}$ is the volume density of the sample we are imaging. The 3D-rotation operator $\boldsymbol{R}_{\theta}(\cdot)$ tilts the volume by an angle $\theta$ around a fixed axis, $\boldsymbol{P}: \mathbbm{R}^{N \times N \times N } \mapsto \mathbbm{R}^{N \times N}$ is the projection operator and  $\boldsymbol{\epsilon}$ is the Gaussian noise which accounts for the low dose electron beam. The tilt $\theta_m$ can only vary between -70 and +70 degrees resulting in a \textit{missing wedge} of measurements. 

This model accounts for the image formation case in which there are no mechanical faults or the sample is not affected by the electron beam. However, this is hardly the case in many applications of cryoET. Accounting for these deformations is crucial to obtain a high resolution reconstruction \cite{mastronarde2006fiducial} and can be taken into account by the following forward model:
\begin{equation}
\label{eq:deformation_model}
     \yb_m = \Db_m(\Pb(\Rb_{\theta_m}(\rhob))) + \boldsymbol{\epsilon}_m,
\end{equation}
where $\Db_m$ models deformations appearing in cryoET. Deformations are composed of rigid deformations due to the sample slightly moving between tilts, and local and non-rigid deformations due to the electron beam interacting with the sample. In this paper, we consider independent deformations on the tilt-series.


\subsection{Related work}

AreTomo \cite{zheng2022aretomo} is one such approach where patches from images are tracked to account for and correct deformations. Such methods that use patch-based tracking can be useful in many applications but cannot track small-scale deformations that can occur within the patch nor deformations that appear between patches. Our approach is similar to AreTomo with the difference that we track all pixels simultaneously.

Neural fields (or implicit neural networks) represent continuous signals as maps from coordinates to function values \cite{mildenhall2021nerf}. 
Such models have been used extensively in inverse rendering problems \cite{sitzmann2020implicit}. 
Nerfies \cite{park2021nerfies} was proposed in this problem to capture the deformations that vary continuously between scenes. 
Lately, it has also been successfully applied to model the volume density in cryoEM \cite{zhong2021cryodrgn,levy2022cryoai}, obtaining accurate reconstructions even with strong structural heterogeneity and noise. 
This was extended to Scanning Transmission Electron Microscopy \cite{kniesel2022clean}. 
In this paper, we propose to model both the unknown volume and the unknown deformation using neural fields in cryoET. Contrary to \cite{park2021nerfies}, we capture deformations that change randomly between images of the tilt-series.

Recently, \cite{debarnot2022joint} showed that global deformations (shear, rotations, and shifts) in the images can be recovered by appropriately modeling them and estimating the deformation parameters.
In this paper, we employ a similar approach but we model directly the unknown volume using an implicit neural network. We also consider a wider class of deformations.

\subsection{Contributions}
We propose an unsupervised method to jointly reconstruct the volume and the deformation from noisy and deformed tilt-series in cryoET. 
We extend previous work \cite{debarnot2022joint} by considering a rich class of deformation that accurately model cryoET acquisitions. Moreover, our original framework allows a significant gain in computation time and resources, while offering the possibility to be paired with sophisticated priors in order to compensate for the missing wedge.
\def\sc{0.16}
\begin{figure*}
	\centering
    \includegraphics[width=\sc\linewidth,trim={2cm 1cm 2cm 0cm},clip]{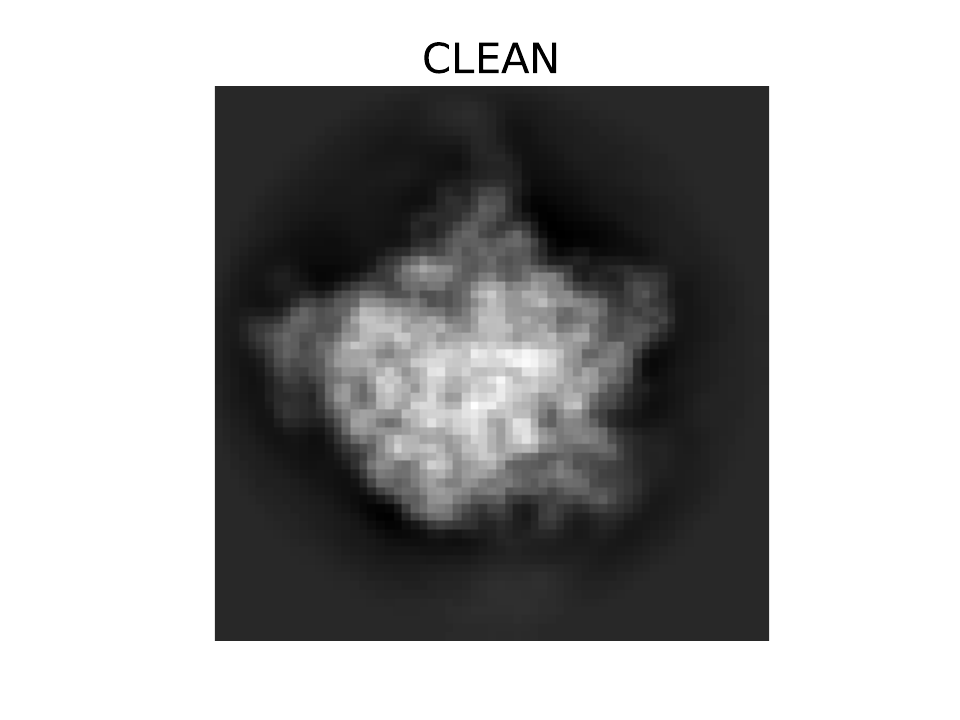}
    \includegraphics[width=\sc\linewidth,trim={2cm 1cm 2cm 0cm},clip]{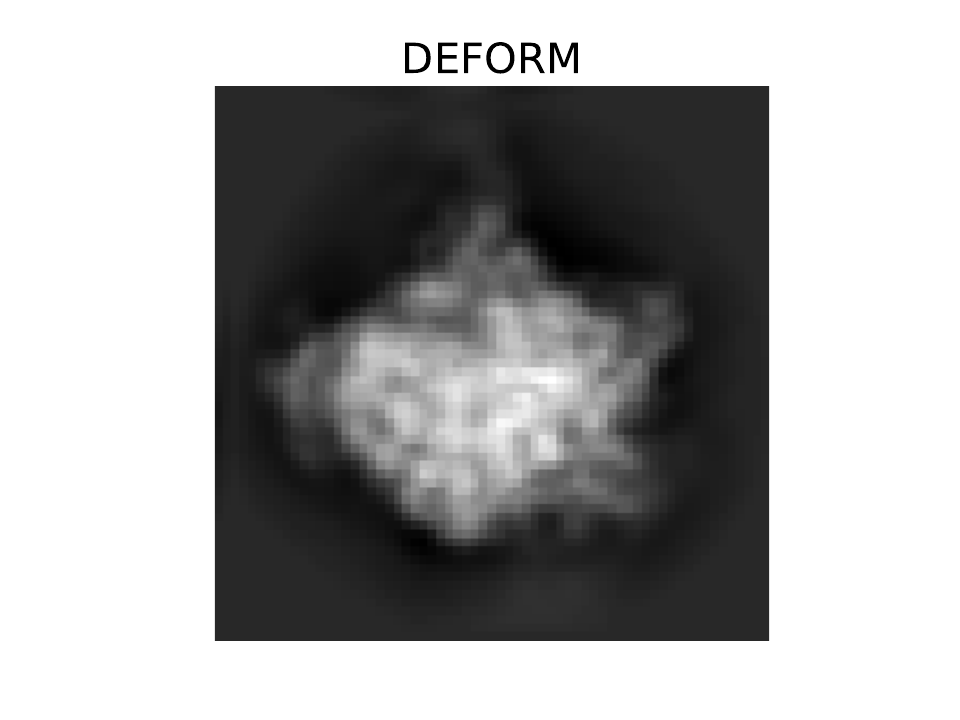}
    \includegraphics[width=\sc\linewidth,trim={2cm 1cm 2cm 0cm},clip]{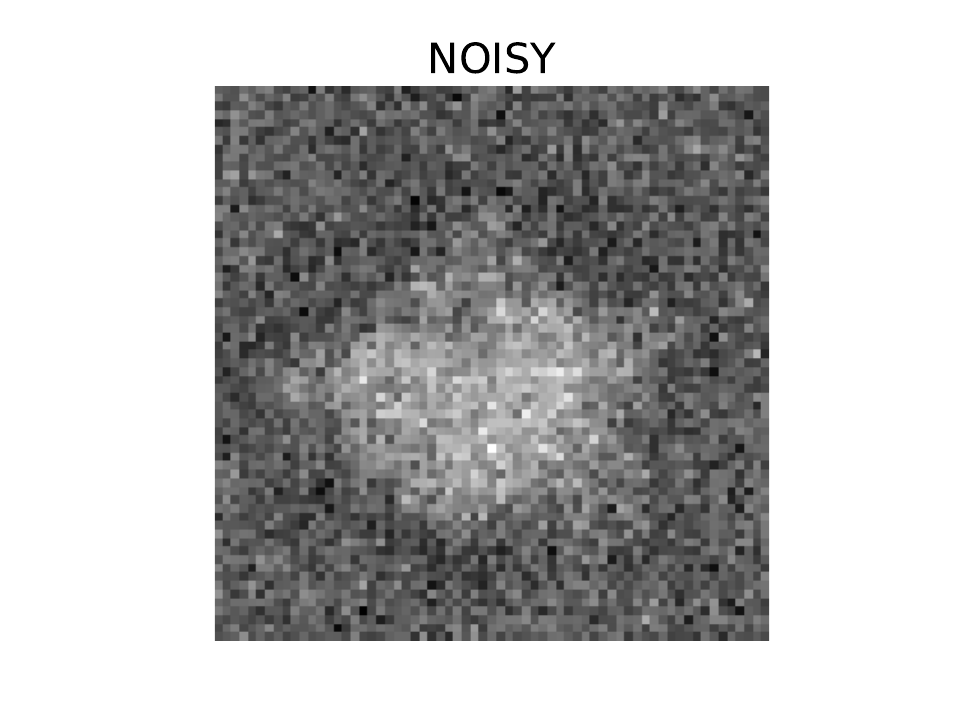}
    \includegraphics[width=\sc\linewidth,trim={2cm 1cm 2cm 0cm},clip]{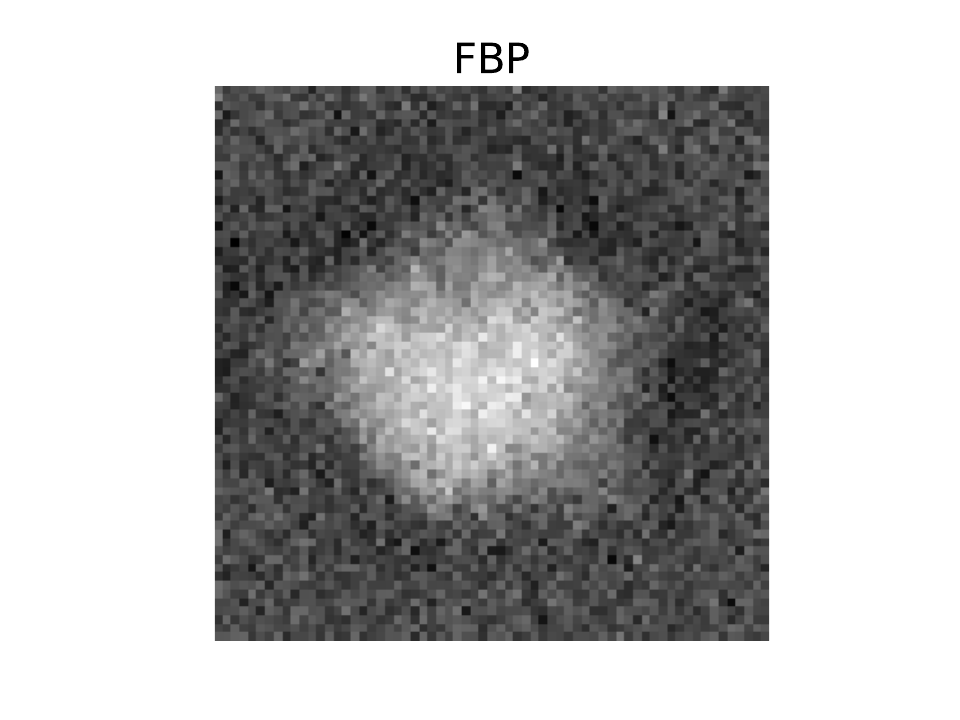}
    \includegraphics[width=\sc\linewidth,trim={2cm 1cm 2cm 0cm},clip]{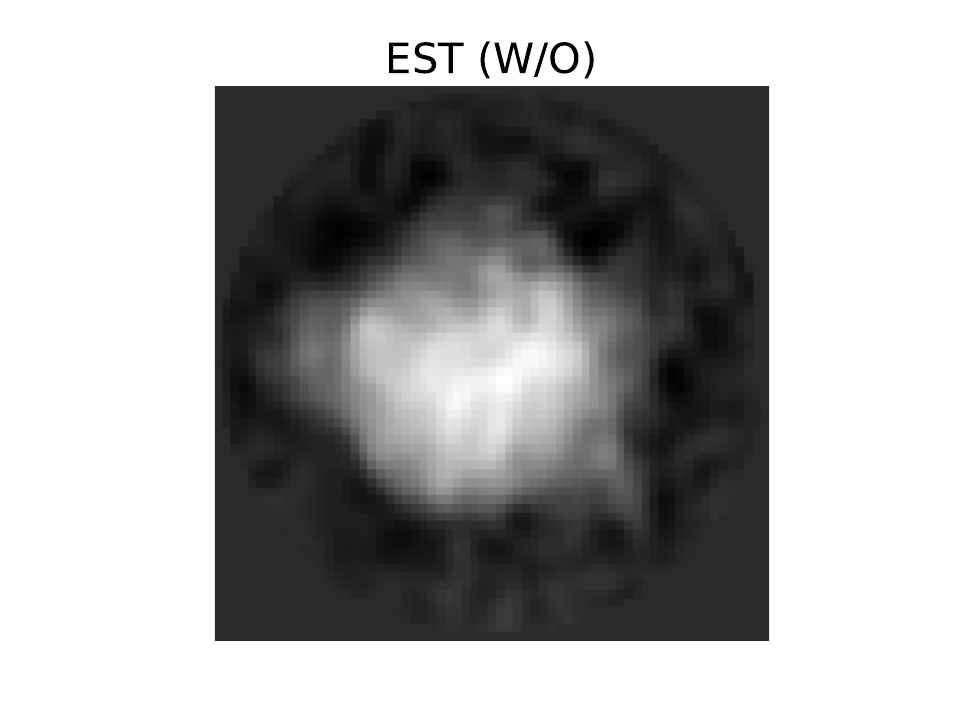}
    \includegraphics[width=\sc\linewidth,trim={2cm 1cm 2cm 0cm},clip]{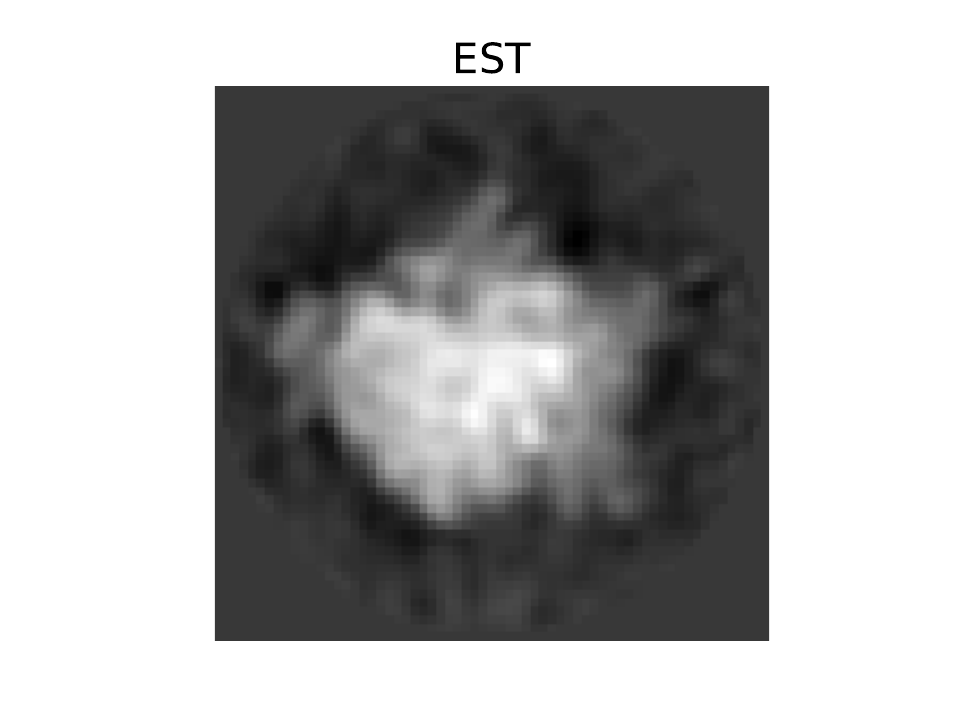}
	\caption{Projections corresponding to the central viewing direction at angle 0 degree. From left to right, projection that is: non-deformed and noise-free, deformed and noise free, noisy and deformed (observation), estimated by the FBP, estimated while not accounting for the deformations (EST W/O), estimated with our method (EST).}
	\label{fig:projections}
 
\vspace{-0.3cm}
\end{figure*}
\section{Methods}
We approximate the volume $\rhob$ by an implicit representation $ \Vb(\psib): \R^3 \to \R,$ parameterized by $\psib$. We use an MLP (Multi-Layer Perceptron) with Fourier feature as positional encoding.
Using an implicit representation of the volume allows fast computation of the formation model \eqref{eq:formation_model} using discretization. The same benefit holds for any deformation acting on the coordinate space.

We approximate the deformations $\Db_m$ as:
\begin{equation*}
  \Db(\phib_m) \eqdef \Lb_m(\gammab_m) \Sb(\taub_m) \Rb_{2D}(\alpha_m)
\end{equation*}

where $\Rb_{2D}(\alpha)$ is a 2D rotation operator by angle $\alpha$, $\Sb(\taub)$ is a shift operator with displacement $\taub = (\tau_x, \tau_y)$ and $\Lb(\gammab): \R^2 \to \R^2$ is an implicit representation, parameterized by $\gammab$, which models local deformations
\begin{align*}
    \left(\Lb(\gammab) v\right)(\xb) = v(\xb+\lb(\gammab)(\xb) ), \forall v\in\Cc^0(\R^2), \forall \xb\in\R^2,
\end{align*}
where $\lb(\gammab):\R^2 \to \R^2$ is a continuous operator in each of its output.
To account for randomness in the local deformations, we use as many networks as there are images, as each network learns the local deformation of its image. 
The implicit representations $\Lb_m$ could also be used to learn the inplane roatation and the shift operators, but we observe better performance when explicitly dissociate these three operators. Similar observation has already been made in Nerfies \cite{park2021nerfies}. 
Unlike Nerfies, we use separate networks to capture local deformations from one image to the other, as we do not have continuous deformations between them. 
We learn the parameters $\{\phib_m = (\gammab_m,\taub_m,\alpha_m)\}_{m=1}^M$ and  $\{\psib_m \}_{m=1}^M$ jointly by minimizing the mean square error between the observations and the estimated tilt-series:
\begin{equation}\label{eq:opt}
    \min_{\{\phib_m\}_{m=1}^M, \{\psib_m\}_{m=1}^M}  \sum_{m=1}^M\|\yb_m - \boldsymbol{D}(\phib_m)(\boldsymbol{P}(\boldsymbol{R}_{\theta_m}(\boldsymbol{\Vb(\psib_m)})) \|_2^2
\end{equation}
Notice that the above formulation allows mini-batch optimization in the pixel space contrary to \cite{debarnot2022joint}.

 \section{Experiment and Results}
We use the voxelized  M. pneumoniae cell \cite{tegunov2021multi} (dataset DOI on EMPIAR 10.6019/EMPIAR-10499) model to test the recovery capacity of the model when there are deformations present. We simulate the cryoET observation by projecting the volume after rotation along a fixed axis at $M=90$ angles uniformly spaced between $-70$ degrees to $70$ degrees. 
We define the signal to noise ration (SNR) between two signals $\sb, \sb_{\text{true}}\in\R^N$ as 
\begin{equation*}
\SNR(\sb, \sb_{\text{true}}) = 10\log_{10}\left(\frac{\var(\sb)}{\var(\sb_{\text{true}})}\right). 
\end{equation*}
We add Gaussian noise so that the noisy projections have an SNR of 0dB.
In the simulations, the true unknown deformations $\Db_m$ are the composition of a smooth random diffeomorphism  as defined in \cite{ronneberger2015u}, a random shift with maximum $\pm 10\%$ of the image size, and a random in-plane rotation with angle between $\pm 10$ degrees.

In the following, we compare our approach that involves estimating deformations (EST) with our approach where we  simply fit an implicit neural network to estimate the volume but without accounting for the deformation (EST W/O). We also compare with the filtered-back projection (FBP) \cite{harauz1986exact}, a standard reconstruction algorithm.

\begin{figure}
	\includegraphics[width=0.48\linewidth]{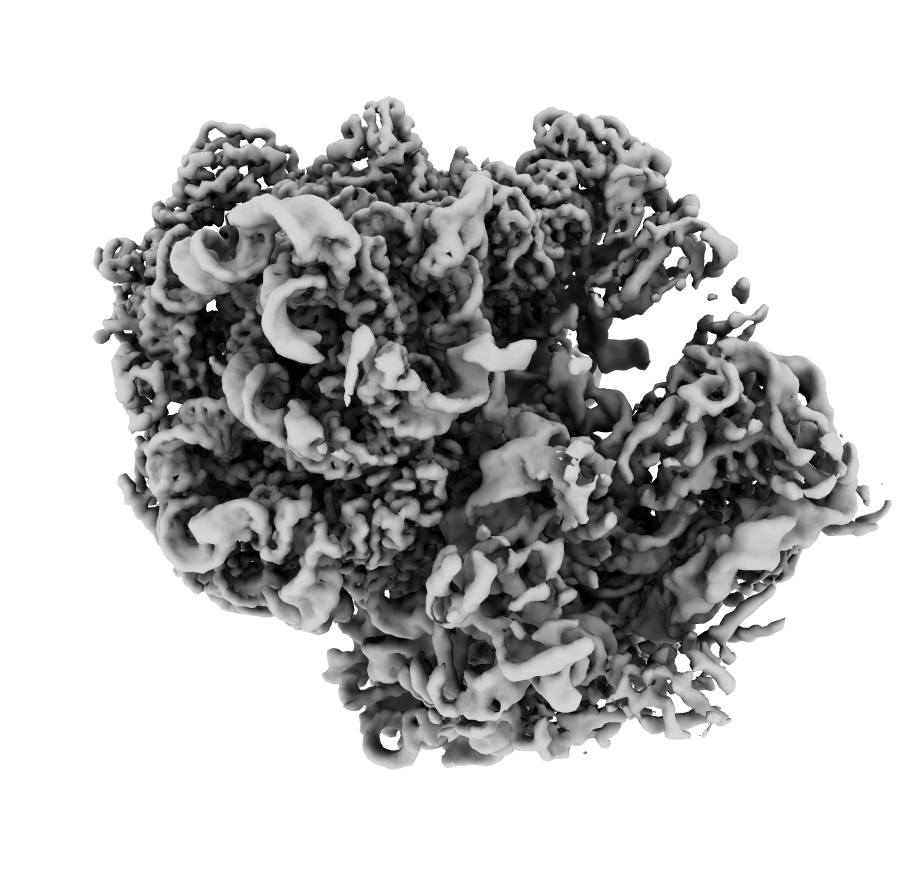}
	\includegraphics[width=0.48\linewidth]{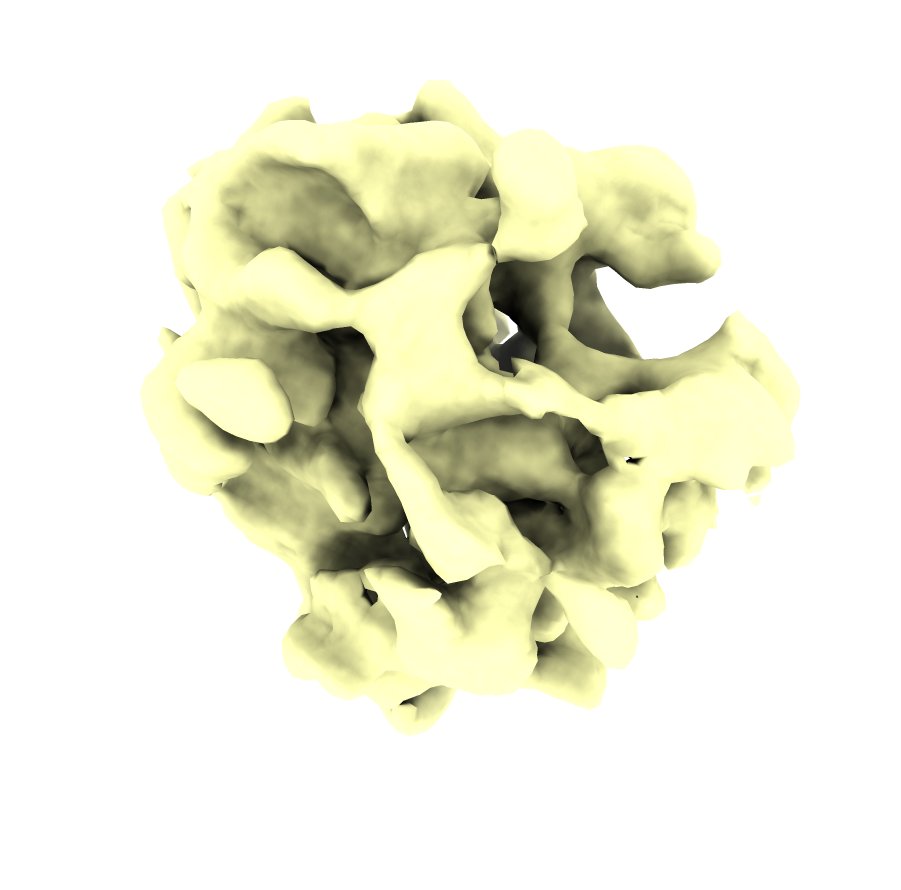}
	\caption{Snapshot of the true (left) and estimated (right) volume.}
	\label{fig:volume}
\end{figure}

\begin{figure}
\vspace{-0.6cm}
	\centering
	\includegraphics[width=0.6\linewidth]{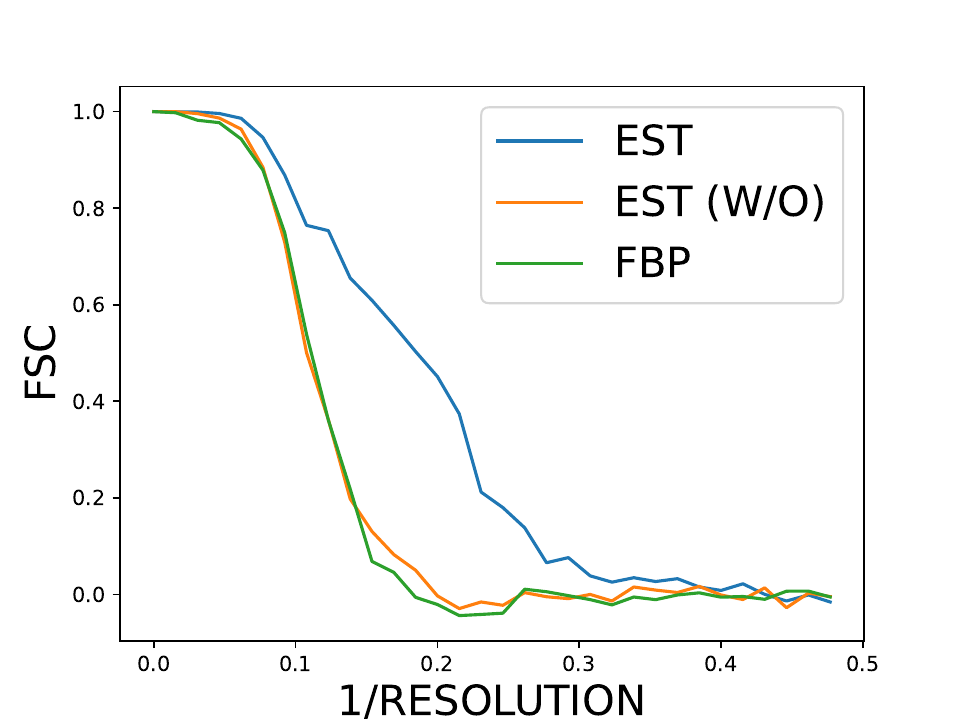}
    \caption{FSC between three different estimations and the true volume. }
	\label{fig:fsc}
 \vspace{-0.5cm}
\end{figure}

We display a snapshot of the true and estimated volumes in Figure \ref{fig:volume}.
We asses the quality of the reconstruction using the Fourier Shell correlation (FSC) in Figure \ref{fig:fsc}. The FSC indicates the correlation between the frequencies of the estimated volume and the original. This is a standard metric used to compare recovered volumes in practice \cite{harauz1986exact}. 
Figure \ref{fig:fsc} shows that estimating the deformation allows a significant increase in the resolution. Notice that all three methods don't compensate for the missing wedge, and we expect a higher gain when incorporating additional prior in Problem \eqref{eq:opt}.

We asses the quality of the estimation of deformation in Table \ref{tab:SNR}. We display the average error on the global deformations (shift and in-plane rotation). We also report the average SNR of the noise-free and non-deformed projections between the true and estimated volume. While the deformation parameters are well estimated, the gain of SNR in the projections also indicates that the volume is better estimated than when no deformation is estimated.

\begin{table}
 \begin{center}
\begin{tabular}{ |c|c|c|c| } 
\hline
 & EST & EST (W/O) & FBP  \\ 
\hline
 Shifts (pixel) & 0.7396 & 3.3863 & 3.3863 \\ 
 In-plane rotation (degree)  & 1.5837 & 5.067 & 5.067 \\ 
local deformation (pixel) & 1.284 & 1.282 &  1.282 \\ 
SNR of projections (dB)  & 14.962 & 11.999 & -6.873 \\ 
 \hline
 \end{tabular}
 \caption{Estimation quality of the parameters at SNR 0dB.}
\label{tab:SNR}
\end{center}
\vspace{-0.5cm}
\end{table}

\section{Conclusion}
We show that pixel-level deformations, widely present in many cryoET applications, can be mitigated using a small set of parameters and networks. Along with this, we use an implicit network to represent the undistorted volume and jointly learn to recover it even in the presence of unknown deformations. We see improvements in the reconstructions. The current setup relies on the inductive bias of the implicit network to accurately recover volume. However, our framework is flexible enough to allow the use of advanced prior, helping to compensate for the missing wedge.

\bibliographystyle{IEEEtran}
\bibliography{references}

\begin{thebibliography}{10}
\providecommand{\url}[1]{#1}
\csname url@samestyle\endcsname
\providecommand{\newblock}{\relax}
\providecommand{\bibinfo}[2]{#2}
\providecommand{\BIBentrySTDinterwordspacing}{\spaceskip=0pt\relax}
\providecommand{\BIBentryALTinterwordstretchfactor}{4}
\providecommand{\BIBentryALTinterwordspacing}{\spaceskip=\fontdimen2\font plus
\BIBentryALTinterwordstretchfactor\fontdimen3\font minus
  \fontdimen4\font\relax}
\providecommand{\BIBforeignlanguage}[2]{{%
\expandafter\ifx\csname l@#1\endcsname\relax
\typeout{** WARNING: IEEEtran.bst: No hyphenation pattern has been}%
\typeout{** loaded for the language `#1'. Using the pattern for}%
\typeout{** the default language instead.}%
\else
\language=\csname l@#1\endcsname
\fi
#2}}
\providecommand{\BIBdecl}{\relax}
\BIBdecl

\bibitem{doerr2017cryo}
A.~Doerr, ``Cryo-electron tomography,'' \emph{Nature Methods}, vol.~14, no.~1,
  pp. 34--34, 2017.

\bibitem{mastronarde2006fiducial}
D.~N. Mastronarde, ``Fiducial marker and hybrid alignment methods for
  single-and double-axis tomography,'' \emph{Electron tomography: methods for
  three-dimensional visualization of structures in the cell}, pp. 163--185,
  2006.

\bibitem{zheng2022aretomo}
S.~Zheng, G.~Wolff, G.~Greenan, Z.~Chen, F.~G. Faas, M.~B{\'a}rcena, A.~J.
  Koster, Y.~Cheng, and D.~A. Agard, ``Aretomo: An integrated software package
  for automated marker-free, motion-corrected cryo-electron tomographic
  alignment and reconstruction,'' \emph{Journal of Structural Biology: X},
  vol.~6, p. 100068, 2022.

\bibitem{mildenhall2021nerf}
B.~Mildenhall, P.~P. Srinivasan, M.~Tancik, J.~T. Barron, R.~Ramamoorthi, and
  R.~Ng, ``Nerf: Representing scenes as neural radiance fields for view
  synthesis,'' \emph{Communications of the ACM}, vol.~65, no.~1, pp. 99--106,
  2021.

\bibitem{sitzmann2020implicit}
V.~Sitzmann, J.~Martel, A.~Bergman, D.~Lindell, and G.~Wetzstein, ``Implicit
  neural representations with periodic activation functions,'' \emph{Advances
  in Neural Information Processing Systems}, vol.~33, pp. 7462--7473, 2020.

\bibitem{park2021nerfies}
K.~Park, U.~Sinha, J.~T. Barron, S.~Bouaziz, D.~B. Goldman, S.~M. Seitz, and
  R.~Martin-Brualla, ``Nerfies: Deformable neural radiance fields,'' in
  \emph{Proceedings of the IEEE/CVF International Conference on Computer
  Vision}, 2021, pp. 5865--5874.

\bibitem{zhong2021cryodrgn}
E.~D. Zhong, T.~Bepler, B.~Berger, and J.~H. Davis, ``Cryodrgn: reconstruction
  of heterogeneous cryo-em structures using neural networks,'' \emph{Nature
  methods}, vol.~18, no.~2, pp. 176--185, 2021.

\bibitem{levy2022cryoai}
A.~Levy, F.~Poitevin, J.~Martel, Y.~Nashed, A.~Peck, N.~Miolane, D.~Ratner,
  M.~Dunne, and G.~Wetzstein, ``Cryoai: Amortized inference of poses for ab
  initio reconstruction of 3d molecular volumes from real cryo-em images,'' in
  \emph{Computer Vision--ECCV 2022: 17th European Conference, Tel Aviv, Israel,
  October 23--27, 2022, Proceedings, Part XXI}.\hskip 1em plus 0.5em minus
  0.4em\relax Springer, 2022, pp. 540--557.

\bibitem{kniesel2022clean}
H.~Kniesel, T.~Ropinski, T.~Bergner, K.~S. Devan, C.~Read, P.~Walther,
  T.~Ritschel, and P.~Hermosilla, ``Clean implicit 3d structure from noisy 2d
  stem images,'' in \emph{Proceedings of the IEEE/CVF Conference on Computer
  Vision and Pattern Recognition}, 2022, pp. 20\,762--20\,772.

\bibitem{debarnot2022joint}
V.~Debarnot, S.~Gupta, K.~Kothari, and I.~Dokmanic, ``Joint cryo-et alignment
  and reconstruction with neural deformation fields,'' \emph{arXiv preprint
  arXiv:2211.14534}, 2022.

\bibitem{tegunov2021multi}
D.~Tegunov, L.~Xue, C.~Dienemann, P.~Cramer, and J.~Mahamid, ``Multi-particle
  cryo-em refinement with m visualizes ribosome-antibiotic complex at 3.5 {\aa}
  in cells,'' \emph{Nature Methods}, vol.~18, no.~2, pp. 186--193, 2021.

\bibitem{ronneberger2015u}
O.~Ronneberger, P.~Fischer, and T.~Brox, ``U-net: Convolutional networks for
  biomedical image segmentation,'' in \emph{Medical Image Computing and
  Computer-Assisted Intervention--MICCAI 2015: 18th International Conference,
  Munich, Germany, October 5-9, 2015, Proceedings, Part III 18}.\hskip 1em plus
  0.5em minus 0.4em\relax Springer, 2015, pp. 234--241.

\bibitem{harauz1986exact}
G.~Harauz and M.~van Heel, ``Exact filters for general geometry three
  dimensional reconstruction.'' \emph{Optik.}, vol.~73, no.~4, pp. 146--156,
  1986.

\end{thebibliography}
\end{document}